\documentclass[pss]{wiley2sp}
\usepackage{amsmath}
\usepackage{bm}
\usepackage{w-greek}
\usepackage{graphicx}
\usepackage{dcolumn}
\usepackage{bm}
\usepackage{rotating}
\usepackage[numbers,square]{natbib} 
\newcommand{\YCS}{YbCo$_{2}$Si$_{2}$}
\newcommand{\YRS}{YbRh$_{2}$Si$_{2}$}
\newcommand{\YCSx}{Yb(Rh$_{1-x}$Co$_x$)$_2$Si$_2$}
\newcommand{\Ir}{Yb(Rh$_{0.94}$Ir$_{0.06}$)$_{2}$Si$_{2}$}
\newcommand{\Co}{Yb(Rh$_{0.93}$Co$_{0.07}$)$_{2}$Si$_{2}$}

\begin{document}
\title{Magnetization study of the energy scales in \YRS\,under chemical pressure}
\titlerunning{Magnetization study of the energy scales in \YRS\,}
%
\author{%
  Manuel Brando\textsuperscript{\Ast,\textsf{\bfseries 1}},
  Luis Pedrero\textsuperscript{\textsf{\bfseries 1}},
  Tanja Westerkamp\textsuperscript{\textsf{\bfseries 1}},
  Cornelius Krellner\textsuperscript{\textsf{\bfseries 1,2}},
  Philipp Gegenwart\textsuperscript{\textsf{\bfseries 1,3}},
  Christoph Geibel\textsuperscript{\textsf{\bfseries 1}},
  and Frank Steglich\textsuperscript{\textsf{\bfseries 1}}}
\authorrunning{M. Brando et al.}
\mail{e-mail
    \textsf{brando@cpfs.mpg.de}, Phone: +49 351 4646 2324, Fax: +49 351 4646 2360,
    Web: \textsf{www.cpfs.mpg.de}}
\institute{%
  \textsuperscript{1}\,Max Planck Institute for
Chemical Physics of Solids, N\"othnitzer Strasse 40, D-01187 Dresden, Germany.\\
  \textsuperscript{2}\,Institute of Physics, Goethe University Frankfurt, Max-von-Laue-Strasse 1, 60438 Frankfurt am Main, Germany\\
  \textsuperscript{3}\,I. Physikalisches Institut, Georg-August-Universit\"at, 37077 G\"ottingen, Germany}
\received{XXXX, revised XXXX, accepted XXXX}
\published{XXXX}

%
\abstract{
\abstcol{
We present a systematic study of the magnetization in \YRS\,under slightly negative (6\% Ir substitution) and positive (7\% Co substitution) chemical pressure. We show how the critical field $H_{0}$, associated with the high-field Lifshitz transitions, is shifted to lower (higher) values with Co (Ir) substitution. The critical field $H_{\mathrm{N}}$, which identifies the boundary line of the antiferromagnetic (AFM) phase $T_{\mathrm{N}}(H)$ increases with positive pressure and it approaches zero with 6\% Ir substitution. On the other side, the crossover field $H^{*}$, associated with the energy scale $T^{*}(H)$ where a reconstruction of the Fermi surface has been observed, is not much influenced by the chemical substitution.}{
Following the analysis proposed in Refs.\,\cite{Paschen2004,Gegenwart2007,Friedemann2009,Tokiwa2009a} we have fitted the quantity $\tilde{M}(H)=M+(dM/dH)H$ with a crossover function to indentify $H^{*}$. The $T^{*}(H)$ line follows an almost linear $H$-dependence at sufficiently high fields outside the AFM phase, but it deviates from linearity at $T \le T_{\mathrm{N}}(0)$ and in \Co\,it changes slope clearly inside the AFM phase. Moreover, the FWHM of the fit function depends linearly on temperature outside the phase, but remains constant inside, suggesting either that such an analysis is valid only for $T \ge T_{\mathrm{N}}(0)$ or that the Fermi surface changes continuously at $T = 0$ inside the AFM phase.}}
\maketitle
\noindent
The general understanding of quantum critical points (QCPs) is based on the concept of a single energy scale that fades continuously for $T \rightarrow 0$. The conventional theoretical approach associates with this energy scale an order parameter that is defined finite inside a region of the phase diagram and zero elsewhere~\cite{Hertz1976,Millis1993,Moriya1995}. The corresponding phase transition at $T = 0$ is a quantum phase transition (QPT). If this region is separated from the rest by a second-order phase transition line, a QCP exists at the QPT. In materials with magnetic phase transitions, the energy scale is usually considered to be the ordering transition temperature - e.g., in the case of antiferromagnetic (AFM) systems, this is the N\'eel temperature $T_{\mathrm{N}}$ - and the order parameter is the staggered magnetization~\cite{Loehneysen2007,Gegenwart2008}. In metallic systems the magnetic order can be of the spin-density-wave (SDW) type and the same electrons which form the Fermi surface are involved in the QPT. Prominent examples of quantum critical systems are heavy-fermion compounds because of the small energy scale associated with the hybridization between nearly localized \textit{f}-electrons with the conduction electrons. Here, the QPT separates a paramagnetic (PM) heavy Fermi liquid (FL) from an AFM metal. In these systems there are two principal energy scales: $k_{\mathrm{B}}T_{\mathrm{K}}$ and $k_{B}T_{\mathrm{RKKY}}$ which derive from the respective interactions, the Kondo and the RKKY. $T_{\mathrm{K}}$ defines the temperature at which the localized $f$-electrons start to hybridize with the itinerant $d$-electrons to form a larger and heavier Fermi surface, $T_{\mathrm{RKKY}}$ is a measure of the inter-site exchange magnetic coupling. The interplay between these energy scales determines the magnetic ordering temperature $T_{\mathrm{N}}$ and characterizes the QCP~\cite{Doniach1977}. In real systems, however, the situation can be rather more complex, due to the presence of multiple energy scales that can get involved in the QPT~\cite{Gegenwart2007}. In addition, many materials show more than just a single magnetic phase transition. Therefore, experimental studies of quantum criticality become quite demanding but, on the other hand, promising for the discovery of novel correlated phases of condensed matter.

A prototypical example of such a complex system is the tetragonal \YRS, which is particularly suitable for studying QPTs~\cite{Trovarelli2000b,Gegenwart2002,Custers2003}. In fact, this compound has a large $T_{\mathrm{K}} \approx 25$\,K and a very small $T_{\mathrm{N}} = 72$\,mK that can be suppressed by a magnetic field $\mu_{0}H_{\mathrm{N}} = 60$\,mT ($H \perp c$, with $c$ being the magnetically hard axis) or chemical negative pressure ($P \approx -0.25$\,GPa)~\cite{Custers2003,Mederle2002,Macovei2008a}. Three other intriguing features have recently been detected: (i) Another sharp phase transition at $T_{\mathrm{L}} = 2.2$\,mK~\cite{Schuberth2009}, (ii) a kink in the magnetization at $H_{0} \approx 10$\,T~\cite{Tokiwa2005,Gegenwart2006}, and (iii) a crossover energy scale $T^{*}(H)$~\cite{Paschen2004,Gegenwart2007}. 
\begin{figure}[b]
\begin{center}
\includegraphics[width=0.65\columnwidth,angle=-90]{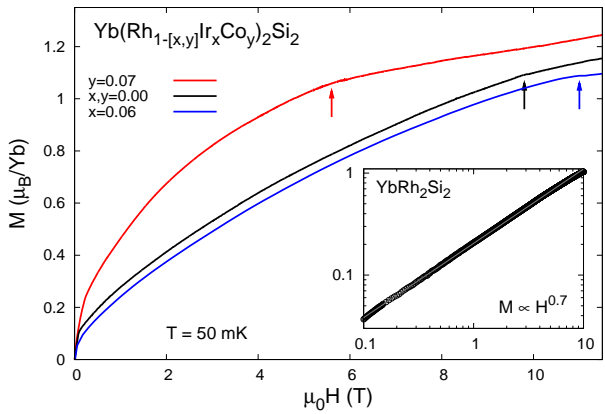}
\end{center}
\caption{Field-dependent magnetization curves for \YRS, \Ir\,and \Co\, samples. The upward arrows ($\uparrow$) indicate the fields $H_{0}$ estimated by the inflection point of $dM(H)/dH$~\cite{Tokiwa2005}. Inset: Log-log plot of the magnetization at 50\,mK of \YRS\,for fields extending from 0.1 to 10\,T, indicating that $M \propto H^{0.7}$.}
\label{fig1}
\end{figure}
The origin of the low-$T$ transition is still unclear, but the comparison with the isoelectronic analogue \YCS\,\cite{Pedrero2010a,Pedrero2011} and the evolution of $T_{\mathrm{L}}$ observed in the series \YCSx\,(Co substitution corresponds to positive pressure) might suggest a second, possibly first-order, AFM transition~\cite{Klingner2009,Klingner2011,Klingner2011a}. The feature at $H_{0}$ has been interpreted as 
field-induced suppression of the HF state, as hydrostatic pressure experiments have revealed a clear correspondence between $H_{0}$ and the Kondo scale $T_{\mathrm{K}}$~\cite{Tokiwa2005,Gegenwart2006}. Accurate de-Haas-van-Alphen experiments could show that the Fermi surface smoothly changes at $H_{0}$ suggesting a Lifshitz-like type of transition~\cite{Rourke2008}. Meanwhile, thermopower experiments and renormalized band structure calculations have undoubtly demonstrated that the anomaly at $H_{0}$ is caused by the field-induced shift of a van-Hove singularity (in the quasiparticle density of states) through the Fermi level, causing two consecutive Lifshitz transitions~\cite{Zwicknagl2011,Pfau2012}. Finally, there is the crossover line $T^{*}(H)$ which has been found in measurements of the Hall-effect~\cite{Paschen2004} and various other thermodynamic properties~\cite{Gegenwart2007}. The full width at half maximum (FWHM) of these crossovers displays a linear temperature dependence~\cite{Friedemann2009,Friedemann2010}. This suggests a step-like change of the Hall coefficient at $T = 0$ implying a Fermi surface reconstruction.
These findings have corroborated a series of previous theoretical proposals which considered the Fermi surface collapse due to the critical breakdown of the Kondo screening effect at the field-induced AFM QCP, including degrees of freedom other than fluctuations of the order parameter~\cite{Coleman2001,Si2001,Senthil2003,Pepin2005}. In addition, it has been shown that the linear dependence of the FWHM on temperature is consistent with the energy over temperature scaling of the quantum-critical single-electron fluctuation spectrum~\cite{Friedemann2010}. However, experimental evidence that the energy scale $T^{*}(H)$ does not change much under applied pressure~\cite{Friedemann2009,Tokiwa2009a}, while the other energy scales $T_{\mathrm{L}}(H)$, $T_{\mathrm{N}}(H)$ and $T_{0}(H)$ are very pressure sensitive~\cite{Mederle2002,Macovei2008a,Klingner2011a}, has reopened the debate on how to interpret the experimental results. Three possibilities are currently considered: (i) The $T^{*}(H)$ line represents a Kondo-destruction Lifshitz transition inside the magnetic phase with a change between two Fermi surfaces which have different topology~\cite{Si2006,Vojta2008,Yamamoto2009,Coleman2010}; (ii) it represents the effect of a Zeeman-induced Lifshitz transition~\cite{Hackl2011,Friedemann2012,Hackl2012}; (iii) recent inelastic neutron scattering experiments associate the electron spin resonance signal~\cite{Sichelschmidt2003} seen in \YRS\,with a field-induced mesoscopic spin resonance, which evolves in field like the $T^{*}(H)$ line~\cite{Stock2012}.

In this article we follow the evolution of the three aforementioned energy scales under the effect of chemical pressure by means of magnetization measurements. Pressure is induced by substituting a small amount of either Co (positive pressure) or Ir (negative pressure) for Rh. The substitution is isoelectronic and small, to avoid the effect of disorder. The single crystals were grown from In flux as described in Ref.~\cite{Krellner2012}. The good quality of the samples as well as the good agreement between chemical and hydrostatic pressure is shown in the Supplementary Information of Ref.~\cite{Friedemann2009}.
\begin{figure}[t]
\begin{center}
\includegraphics[width=0.95\columnwidth,angle=0]{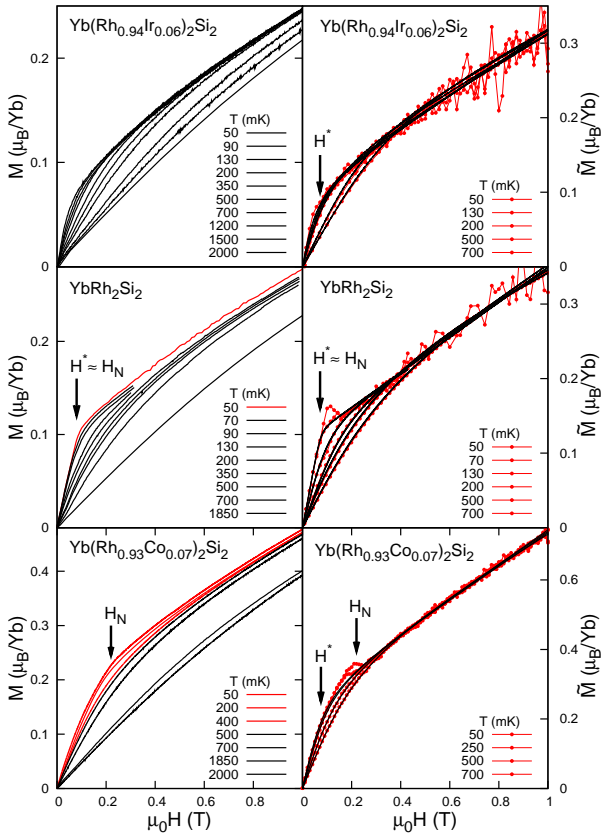}
\end{center}
\caption{Left: Magnetization isotherms for the three single crystals with field $H \perp c$. Red lines indicate measurements at temperatures $T < T_{\mathrm{N}}$. $H_{\mathrm{N}}$ and $H^{*}$ are the fields associated with $T_{\mathrm{N}}$ and $T^{*}$ at 50\,mK. In \YRS\,the two fields almost coincide. Right: $\tilde{M}=M+(dM/dH)H$ vs. $H$ for the same three single crystals. The little humps, visible just above $H_{\mathrm{N}}$, denote the phase transition and their shape is a consequence of how $\tilde{M}$ is calculated.}
\label{fig3}
\end{figure}
Our crystals of \YRS, \Co\,and \Ir\,have a residual resistivity of 0.55, 3, and 7.4\,$\mu\Omega$cm, respectively. In the \Co\,sample two phase transitions have been found at $T_{\mathrm{N}} = 410$\,mK and $T_{\mathrm{L}} = 60$\,mK, whereas in \Ir\, no phase transition has been seen down to 20\,mK. The dc-magnetization $M$ has been measured with a high-resolution Faraday magnetometer, in magnetic fields as high as 12\,T and temperatures down to 50\,mK~\cite{Sakakibara1994}. The field was applied along the magnetic easy plane, i.e., perpendicular to the crystallographic $c$-axis. We show how the critical field $H_{0}$ is shifted to lower (higher) values with Co (Ir) substitution, as expected~\cite{Tokiwa2005}. On the other side, the crossover field $H^{*}$ is not much influenced by the chemical substitution. Performing the analysis proposed in Refs.\,\cite{Paschen2004,Gegenwart2007,Friedemann2009,Tokiwa2009a} we illustrate that $T^{*}(H)$ seems to follow a linear $H$ dependence outside the AFM phase, deviating from linearity at temperatures close to $T_{\mathrm{N}}(0)$, and in \Co\,it changes slope inside the magnetic phase. Moreover, the FWHM of the crossover fit function is linear-in-$T$ outside the AFM phases but constant inside the \Co\,AFM phase, suggesting that either such an analysis is valid only for $T \ge T_{\mathrm{N}}(0)$ or the Fermi surface changes continuously at $T = 0$. 

The first result is illustrated in Fig.\,\ref{fig1}, where the field-dependent magnetization curves for \YRS, \Ir\,and \Co\, samples are shown up to 12\,T. In \YRS\,the magnetization measured at 50\,mK shows two clear kinks at about 0.1 (cf. Fig.\,\ref{fig3}) and 10\,T~\cite{Tokiwa2005}. Inbetween $M$ is proportional to $H^{0.7}$, as demonstrated in the inset of the same figure, possibly reflecting the continuous evolution with magnetic field of the quasiparticle density of states at the Fermi energy~\cite{Tokiwa2005,Gegenwart2006} given as a consequence of the suppression of the local Kondo effect~\cite{Hewson2006,Zwicknagl2011}. The upward arrows ($\uparrow$) indicate the fields $H_{0}$, associated with the high-field Lifshitz transitions~\cite{Tokiwa2005,Steppke2010}. In the \Co\, sample the kink at $H_{0}$ is shifted to about 5.6\,T while in \Ir\, $H_{0}$ is shifted to 11\,T, in agreement with previous pressure studies~\cite{Tokiwa2005}.

We focus now on the energy scale $T^{*}(H)$. The field $H^{*}$, associated with this energy scale, was identified in \YRS\, with the change of slope of the quantity $\tilde{M}(H)=M+(dM/dH)H$, i.e. the kink in $\tilde{M}$ vs. $H$, derived from magnetization isotherms (cf. Fig.~\ref{fig3}, right panels)~\cite{Gegenwart2007,Tokiwa2009a,Gegenwart2008b}. At 50\,mK, the kink in $\tilde{M}(H)$ at about $H^{*} \approx 50$\,mT derives from the kink in $M(H)$ at about 100\,mT which remains sharp even at temperatures higher than $T_{\mathrm{N}} \approx 72$\,mK (cf. curves in middle panels of Fig.~\ref{fig3} at 70 and 90\,mK). Therefore, the kink at $H^{*}$ can not be associated with the critical field $H_{\mathrm{N}}$ of the AFM phase, but matches quite well the crossover field extracted from Hall-effect, magnetoresistivity and ac-susceptibility measurements~\cite{Gegenwart2007,Friedemann2009,Friedemann2010}. Moreover, $H^{*}$ is also visible in \Ir\, where $T_{\mathrm{N}}$ is almost zero. We have performed the analysis proposed in Refs.\,\cite{Paschen2004,Gegenwart2007} for the three single crystals (see Fig.\,\ref{fig3}). In analogy with the Hall-effect signatures observed at $T^{*}(H)$, $M$ vs. $H$ might be fitted with the integral of the following step function: 
\begin{equation}
\label{eq1}
f(H,T)=A_{2}-\frac{A_{2}-A_{1}}{1+(H/H^{*})^{p}}
\end{equation}
\begin{figure}[!ht]
\begin{center}
\includegraphics[width=0.95\columnwidth,angle=0]{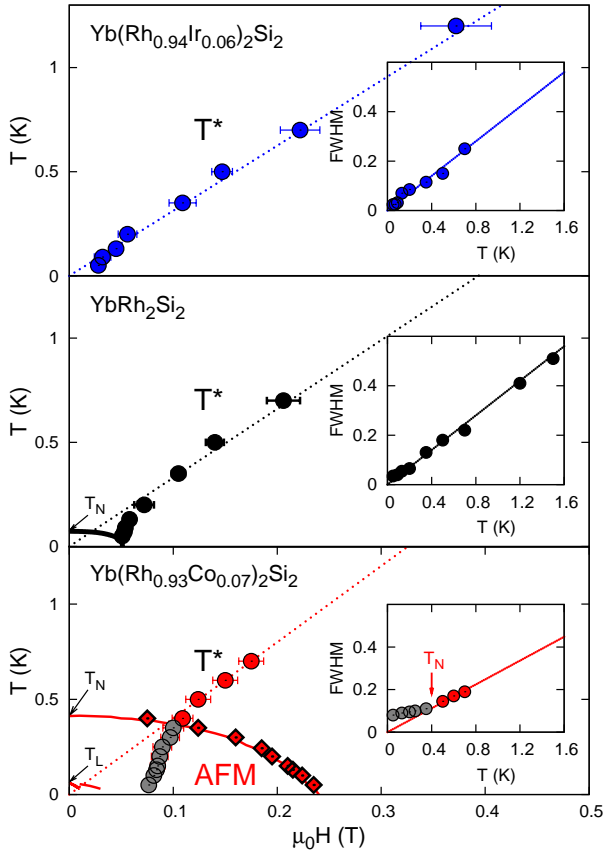}
\end{center}
\caption{Magnetic phase diagrams for the three single crystals with field $H \perp c$. The AFM phase boundary lines has been obtained by $T$- and $H$-dependent ac-susceptibility measurements (black and red solid lines)~\cite{Westerkamp2009}; for \Co, points detected from magnetization isotherms have been included (red diamonds). Circles correspond to the $T^{*}(H)$ line derived by the analysis of $\tilde{M}$ vs. $H$ (gray inside the AFM phase). Dotted lines are a guide to the eye. Inset: FWHM of the fit function (Eq.\,\ref{eq1}). The solid lines are linear fits of the points located above $T_{\mathrm{N}}$.}
\label{fig5}
\end{figure}
where parameters $A_{1}$ and $A_{2}$ denote the linear slope of $M$ vs. $H$ before and after the kink. Since $M$ vs. $H$ is not linear for $H \ge H^{*}$, we have previously used the quantity $\tilde{M}=M+(dM/dH)H$ vs. $H$ which represents the derivative of the magnetic free energy and is almost linear above $\mu_{0}H = 0.5$\,T up to at least 2\,T (right frames of Fig.\,\ref{fig3}). We have performed these fits for all our isotherms as shown in the right panels of Fig.~\ref{fig3}. It is worth mentioning, that, for instance, the fact that at 50\,mK in the stoichiometric crystal the critical fields $H_{\mathrm{N}}$ and $H^{*}$ almost coincide would imply that at temperatures lower that 50\,mK these energy scales might intersect each other (this hypothesis is currently being investigated by measurements of the Hall effect under pressure). For Co concentrations higher than 7\% this analysis clould not be performed anymore because of the high $H_{\mathrm{L}}$ and the stronger curvature of $M$ vs. $H$.
\begin{figure}[!ht]
\begin{center}
\includegraphics[width=0.65\columnwidth,angle=-90]{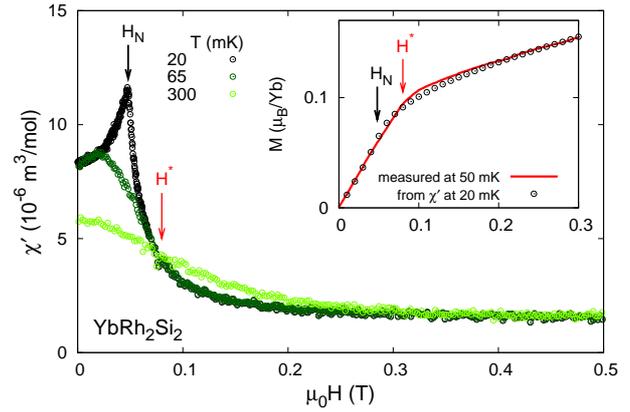}
\end{center}
\caption{Field-dependent ac-susceptibility measurements at temperatures above and below $T_{\mathrm{N}} = 72$\,mK. The peak at 20\,mK indicates the critical field $H_{\mathrm{N}} = 50$\,mT. Inset: Measured magnetization at 50\,mK plotted with the magnetization calculated out of the $\chi'(H)$ data at 20\,mK. The red arrow indicates the position of the kink in $M(H)$ which corresponds to $H^{*}$ where $d\chi'(T)/dT = 0$.}
\label{fig2}
\end{figure}
In the right frames of Fig.\,\ref{fig3} the black lines are the fit to the data performed by integrating equation\,\ref{eq1}. The little humps are a consequence of the weak kinks in the magnetization isotherms due to the transition at $H_{\mathrm{N}}$ from the AFM phase into the PM phase. At these kinks $dM(H)/dH$ decreases slightly and $\tilde{M}(H)$ shows a drop which is small when compared to the main magnetization signal. These humps are not considered during the fit procedures. This can be seen in the right panel of Fig.~\ref{fig3} where $\tilde{M}(H)$ vs. $H$ is plotted for \YRS: The fit function for the data at 70\,mK which show no hump since $T \approx T_{\mathrm{N}}$ lies on the top of the fit function for the data at 50\,mK which show a clear hump at $H_{\mathrm{N}}$. The same is valid for \Co\,where $\tilde{M}(H)$ displays a distinct curvature already at fields $H < H_{\mathrm{N}}$ (see lower right panel of Fig.~\ref{fig3}). The fact that $M(H)$ vs. $H$ does not change much when the temperature is lowered below $T_{\mathrm{N}}$ is reflected in the almost constant position of the $T^{*}(H)$ line inside the AFM phase (gray points in Fig.~\ref{fig5}), as observed before~\cite{Friedemann2009}. 

The results of such fits are summarized in the phase diagrams of Fig.\,\ref{fig5}: The crossover field $H^{*}$, associated with the energy scale $T^{*}(H)$, is not much influenced by the chemical substitution when compared to the substantial change of the AFM ordered phase, i.e. the enhancement of $T_{\mathrm{N}}$ and $H_{\mathrm{N}}$. $T^{*}(H)$ seems to follow a linear $H$ dependence outside the AFM phase, it deviates from linearity at $T < T_{\mathrm{N}}(0)$ and in the \Co\,sample it changes slope clearly inside the magnetic phase. Correspondingly, the FWHM of the crossover fit function depends linearly on temperature at sufficiently high magnetic fields outside the AFM phase in agreement with Ref.~\cite{Friedemann2010}, but remains constant inside, suggesting either that such an analysis is valid only for $T \ge T_{\mathrm{N}}(0)$, where it is not influenced by the ordered magnetic structure, or that the Fermi surface changes continuously at $T = 0$ inside the magnetic phase. The pronounced change of slope inside the AFM phase of the \Co\,sample was not seen in ac-susceptibility measurements where $H^{*}(0) \approx 0.06$\,T~\cite{Friedemann2009}. Our analysis of the magnetization provides $H^{*}(0) \approx 0.075$\,T.

We discuss now the position of $H^{*}$ with respect to $H_{\mathrm{N}}$ in \YRS. To detect the precise position of $H_{\mathrm{N}}$ we have measured the $H$-dependence of the ac-susceptibility $\chi'$ in the very same sample at temperatures below and above $T_{\mathrm{N}}$. The results are shown in Fig.\,\ref{fig2}: At 20\,mK, $\chi'(H)$ displays a clear peak at $H_{\mathrm{N}} \approx 50$\,mT and then decreases rapidly and continuously. This feature indicates a metamagnetic-like transition from the AFM to a PM state. The kink in $M(H)$ at about 0.1\,T, which is in turn associated with the kink at $H^{*} \approx 0.05$\,T in $\tilde{M}(H)$, is the result of the rapid flattening of $\chi'(H)$ with increasing $H$. No other anomalies are detected. Integrating the curve at 20\,mK and plotting it together with the magnetization measured at 50\,mK (inset of Fig.\,\ref{fig2}), we observe that the two curves match relatively well. From this analysis it can be inferred that $H_{\mathrm{N}} < H^{*}$ in agreement with results from magnetotransport experiments~\cite{Friedemann2010}. Moreover, the red arrow in Fig.~\ref{fig2}, which marks the position of $H^{*}$, points towards the point where the three susceptibility curves cross each other (isosbestic point~\cite{Eckstein2007}). This indicates that $d\chi'(T)/dT = 0$ in agreement with the maximum observed at $T^{*}$ in $\chi'(T)$~\cite{Friedemann2009}. The Maxwell relation $(\partial S / \partial H )_{T} = ( \partial M / \partial T )_{H}$ implies that the field dependence of the entropy $S(H)$ has an inflection point at $H^{*}$ as it was demonstrated in Refs.~\cite{Tokiwa2009a,Tokiwa2009}. Thus, the $T^{*}(H)$ line defines the lines of $d^{2}S/dH^{2} = 0$.

To conclude, we have analyzed magnetization isotherms of single crystals of \YRS, \Ir\,and \Co\,to indentify the position of the Fermi reconstruction crossover line $T^{*}(H)$ in the $H - T$ phase diagrams. We confirm that $T^{*}(H)$ is not much influenced by the Co and Ir isoelectronic substitution and that this line follows the points in the phase diagram where the entropy shows an inflection point. In the phase diagram of \YRS\,this line is definitely located on the right of the AFM phase boundary line at the lowest temperatures. More importantly, in \Co\, the $T^{*}(H)$ line clearly falls inside the AFM phase and the FWHM of the crossover function remains almost constant inside the AFM phase while $T \rightarrow 0$, suggesting either that such an analysis is valid only for $T \ge T_{\mathrm{N}}(0)$ or that the Fermi surface changes continuously at $T = 0$.
\begin{acknowledgement}
We are indebted to S. Friedemann, Q. Si, A. Steppke, Y. Tokiwa, M. Vojta, S. Wirth for motivating discussions. Part of the work was supported by the DFG Research Unit 960 ``Quantum Phase Transitions''.
\end{acknowledgement}
\bibliographystyle{h-physrev}
\bibliography{brando_pps_2012}

\begin{thebibliography}{10}

\bibitem{Paschen2004}
S.~Paschen {\em et~al.},
\newblock Nature {\bf 432}, 881 (2004).

\bibitem{Gegenwart2007}
P.~Gegenwart {\em et~al.},
\newblock Science {\bf 315}, 969 (2007).

\bibitem{Friedemann2009}
S.~Friedemann {\em et~al.},
\newblock Nature Phys. {\bf 5}, 465 (2009).

\bibitem{Tokiwa2009a}
Y.~Tokiwa, P.~Gegenwart, C.~Geibel, and F.~Steglich,
\newblock J. Phys. Soc. Jpn. {\bf 78}, 123708 (2009).

\bibitem{Hertz1976}
J.~A. Hertz,
\newblock Phys. Rev. B {\bf 14}, 1165 (1976).

\bibitem{Millis1993}
A.~J. Millis,
\newblock Phys. Rev. B {\bf 48}, 7183 (1993).

\bibitem{Moriya1995}
T.~Moriya and T.~Takimoto,
\newblock J. Phys. Soc. Jpn. {\bf 64}, 960 (1995).

\bibitem{Loehneysen2007}
H.~v. L\"ohneysen, A.~Rosch, M.~Vojta, and P.~W\"olfle,
\newblock Rev. Mod. Phys. {\bf 79}, 1015 (2007).

\bibitem{Gegenwart2008}
P.~Gegenwart, Q.~Si, and F.~Steglich,
\newblock Nature Phys. {\bf 4}, 186 (2008).

\bibitem{Doniach1977}
S.~Doniach,
\newblock Physica B+C {\bf 91}, 231 (1977).

\bibitem{Trovarelli2000b}
O.~Trovarelli {\em et~al.},
\newblock Phys. Rev. Lett. {\bf 85}, 626 (2000).

\bibitem{Gegenwart2002}
P.~Gegenwart {\em et~al.},
\newblock Phys. Rev. Lett. {\bf 89}, 056402 (2002).

\bibitem{Custers2003}
J.~Custers {\em et~al.},
\newblock Nature {\bf 424}, 524 (2003).

\bibitem{Mederle2002}
S.~Mederle {\em et~al.},
\newblock J. Phys.: Condens. Matter {\bf 14}, 10731 (2002).

\bibitem{Macovei2008a}
M.~E. Macovei, M.~Nicklas, C.~Krellner, C.~Geibel, and F.~Steglich,
\newblock J. Phys.: Condens. Matter {\bf 20}, 505205 (2008).

\bibitem{Schuberth2009}
E.~Schuberth {\em et~al.},
\newblock J. Phys.: Conf. Ser. {\bf 150}, 042178 (2009).

\bibitem{Tokiwa2005}
Y.~Tokiwa {\em et~al.},
\newblock Phys. Rev. Lett. {\bf 94}, 226402 (2005).

\bibitem{Gegenwart2006}
P.~Gegenwart {\em et~al.},
\newblock New J. Phys. {\bf 8}, 171 (2006).

\bibitem{Pedrero2010a}
L.~Pedrero {\em et~al.},
\newblock J. Phys.: Conf. Ser. {\bf 200}, 012157 (2010).

\bibitem{Pedrero2011}
L.~Pedrero {\em et~al.},
\newblock Physical Review B {\bf 84}, 224401 (2011).

\bibitem{Klingner2009}
C.~Klingner,
\newblock Dipl. thesis, Technical University, Dresden, 2009.

\bibitem{Klingner2011}
C.~Klingner, C.~Krellner, M.~Brando, C.~Geibel, and F.~Steglich,
\newblock New Journal of Physics {\bf 13}, 083024 (2011).

\bibitem{Klingner2011a}
C.~Klingner {\em et~al.},
\newblock Physical Review B {\bf 83}, 144405 (2011).

\bibitem{Rourke2008}
P.~M.~C. Rourke {\em et~al.},
\newblock Phys. Rev. Lett. {\bf 101}, 237205 (2008).

\bibitem{Zwicknagl2011}
G.~Zwicknagl,
\newblock Journal of Physics: Condensed Matter {\bf 23}, 094215 (2011).

\bibitem{Pfau2012}
H.~Pfau,
\newblock unpublished .

\bibitem{Friedemann2010}
S.~Friedemann {\em et~al.},
\newblock Proc. Natl. Acad. Sci. {\bf 107}, 14547 (2010).

\bibitem{Coleman2001}
P.~Coleman, C.~P\'epin, Q.~Si, and R.~Ramazashvili,
\newblock J. Phys.: Condens. Matter {\bf 13}, R723 (2001).

\bibitem{Si2001}
Q.~Si, M.~S. Rabello, K.~Ingersent, and J.~L. Smith,
\newblock Nature {\bf 413}, 804 (2001).

\bibitem{Senthil2003}
T.~Senthil, S.~Sachdev, and M.~Vojta,
\newblock Phys. Rev. Lett. {\bf 90}, 216403 (2003).

\bibitem{Pepin2005}
C.~P\'epin,
\newblock Phys. Rev. Lett. {\bf 94}, 066402 (2005).

\bibitem{Si2006}
Q.~Si,
\newblock Physica B {\bf 378-380}, 23 (2006).

\bibitem{Vojta2008}
M.~Vojta,
\newblock Phys. Rev. B {\bf 78}, 125109 (2008).

\bibitem{Yamamoto2009}
S.~J. Yamamoto and Q.~Si,
\newblock Phys. Rev. Lett. {\bf 99}, 016401 (2007).

\bibitem{Coleman2010}
P.~Coleman and A.~H. Nevidomskyy,
\newblock Journal of Low Temperature Physics {\bf 161}, 182 (2010).

\bibitem{Hackl2011}
A.~Hackl and M.~Vojta,
\newblock Phys. Rev. Lett. {\bf 106}, 137002 (2011).

\bibitem{Friedemann2012}
S.~Friedemann {\em et~al.},
\newblock arXiv:1207.0536  (2012).

\bibitem{Hackl2012}
A.~Hackl and M.~Vojta,
\newblock arXiv:1207.1123  (2012).

\bibitem{Sichelschmidt2003}
J.~Sichelschmidt, V.~A. Ivanshin, J.~Ferstl, C.~Geibel, and F.~Steglich,
\newblock Physical Review Letters {\bf 91}, 156401 (2003).

\bibitem{Stock2012}
C.~Stock {\em et~al.},
\newblock Phys. Rev. Lett. {\bf 109}, 127201 (2012).

\bibitem{Krellner2012}
C.~Krellner, S.~Taube, T.~Westerkamp, Z.~Hossain, and C.~Geibel,
\newblock Philosophical Magazine {\bf 92}, 2508 (2012).

\bibitem{Sakakibara1994}
T.~Sakakibara, H.~Mitamura, T.~Tayama, and H.~Amitsuka,
\newblock Jpn. J. Appl. Phys. {\bf 33}, 5067 (1994).

\bibitem{Hewson2006}
A.~C. Hewson, J.~Bauer, and W.~Koller,
\newblock Phys. Rev. B {\bf 73}, 045117 (2006).

\bibitem{Steppke2010}
A.~Steppke {\em et~al.},
\newblock Phys. Status Solidi (b) {\bf 247}, 737 (2010).

\bibitem{Gegenwart2008b}
P.~Gegenwart {\em et~al.},
\newblock Physica B {\bf 403}, 1184 (2008).

\bibitem{Westerkamp2009}
T.~Westerkamp,
\newblock {\em Quantenphasen\"uberg\"ange in den Schwere-Fermionen-Systemen
  Yb(Rh$_{1-x}$M$_x$)$_{2}$Si$_{2}$ und CePd$_{1-x}$Rh$_{x}$},
\newblock PhD thesis, Technische Universit\"at Dresden, 2009.

\bibitem{Eckstein2007}
M.~Eckstein, M.~Kollar, and D.~Vollhardt,
\newblock Journal of Low Temperature Physics {\bf 147}, 279 (2007).

\bibitem{Tokiwa2009}
Y.~Tokiwa, T.~Radu, C.~Geibel, F.~Steglich, and P.~Gegenwart,
\newblock Phys. Rev. Lett. {\bf 102}, 066401 (2009).

\end{thebibliography}
\end{document}